\documentclass[twocolumn,amsmath,aps]{revtex4}
\usepackage{graphicx}

\newcommand{\nl}{\nonumber \\}

\newcommand{\be}{\begin{equation}}
\newcommand{\ee}{\end{equation}}
\newcommand{\bea}{\begin{eqnarray}}
\newcommand{\eea}{\end{eqnarray}}

\newcommand{\Eq}[1]{Eq.\,(\ref{#1})}
\newcommand{\Eqs}[1]{Eqs.\,(\ref{#1})}

\newcommand{\la}{\langle}
\newcommand{\ra}{\rangle}
\newcommand{\dg}{\dagger}
\newcommand{\ep}{\epsilon}

\newcommand{\al}{\alpha}

\newcommand{\ti}{\tilde}
\newcommand{\mb}{\mbox}

\begin{document}
\draft

\title{Quantum master equation approach to quantum transport
       through mesoscopic systems}

\author{ Xin-Qi Li, Jun-Yan Luo and Yong-Gang Yang}
\address{State Key Laboratory for Superlattices and Microstructures,
         Institute of Semiconductors,
         Chinese Academy of Sciences, P.O.~Box 912, Beijing 100083, China}
\author{ Ping Cui and YiJing Yan }
\address{Department of Chemistry, Hong Kong University of Science and
         Technology, Kowloon, Hong Kong }

\date{\today}

\begin{abstract}

For quantum transport through mesoscopic system,
a quantum master equation approach is developed
in terms of compact expressions for the transport current
and the reduced density matrix of the system.
The present work is an extension of Gurvitz's approach
for quantum transport and quantum measurement, namely,
to finite temperature and arbitrary bias voltage.
Our derivation starts from a second-order cummulant
expansion of the tunneling Hamiltonian, then follows
conditional average over the electrode reservoir states.
As a consequence, in the usual weak tunneling regime,
the established formalism is applicable
for a wide range of transport problems.
The validity of the formalism and its convenience
in application are well illustrated by a number of examples. 
\end{abstract}
\maketitle

\section{Introduction}

Quantum transport through mesoscopic nanostructures has revealed
many impressive features associated with a number of unique
natures such as the quantum interferences, discrete levels, and
many-body correlations \cite{Dat95}.
Depending on the specific systems/problems under study,
theoretical formalisms have been developed such as
the Landauer-B\"uttiker theory and the non-equilibrium
Green's function (nGF) approach \cite{Dat95,Hau96}.
However, generally speaking, neither of them implies universal
simplicity in practice, for instance, in treating mesosopic transport
in the presence of many-electron Coulomb interaction and inelastic
scattering with phonons.
In particular, it is even more difficult to describe the
transient processes (i.e. time-dependent transport phenomena).

In some particular cases,
a relatively simpler method being able to address these issues
is the rate equation approach \cite{Gla88,Dav93,Naz93,Gur96a,Gur96b}.
Originally, the ``classical" rate equation is in certain sense
of phenomenological form \cite{Gla88}.
Later efforts include its derivation and {\it quantum} generalization
in the context of the resonant tunneling system,
based on the nGF quantum kinetic theory \cite{Dav93},
as well as its modification to describe quantum coherence which
typically exists in mesoscopic systems \cite{Naz93,Gur96a}.
In particular, a {\it microscopic derivation} starting with the many-particle
wavefunction has been presented by Gurvitz {\it et al} \cite{Gur96b}.
However, an obvious drawback of Gurvitz's approach is its limited validity
conditions (i.e. the large bias voltage and zero temperature), which largely
restricts the applicability.
Also, they were unable to derive a general formula in
a ``system-Hamiltonian-free" form, which means the inconvenience
that one has to proceed derivation from the very beginning
for every specific system in practice.
In this work, we extend Gurvitz's approach to finite temperature
and arbitrary bias voltage, as done in our recent work on
quantum measurement \cite{Gur97,Moz02,Gur03,Rus02,Li04a,Li04b}.
In particular, we will establish compact expressions for the
transport current together with the reduced density matrix,
which can serve as a convenient starting point to study
a variety of mesoscopic transport problems.

The remainder of the paper is organized as follows.
In Sec.\ II, starting with the second-order cummulant
expansion of the tunneling Hamiltonian,
formal expressions for the transport current and the associated
master equation are derived.
Sec.\ III is devoted to a number of examples to illustrate the
application of the established formalism.
Finally, in Sec.\ IV concluding remarks on the approximations adopted
and the connection with nGF approach are presented.
In the Appendix, refinement on the cummulant second-order
approximation is self-consistently made by including the
level broadening effect.

\section{Formalism}

Consider the transport setup schematically shown
in Fig.\ 1 which is described by the following Hamiltonian
\bea\label{H-ms}
H &=& H_S(a_{\mu}^{\dg},a_{\mu})+\sum_{\al=L,R}\sum_{\mu k}
       \ep_{\al\mu k}d^{\dg}_{\al\mu k}d_{\al\mu k}  \nl
  & &  + \sum_{\al=L,R}\sum_{\mu k}(t_{\al\mu k}a^{\dg}_{\mu}
       d_{\al\mu k}+\rm{H.c.}) .
\eea
$H_S$ is  the (mesoscopic) system Hamiltonian,
which can be rather general (e.g. including many-body interaction).
$a^{\dg}_{\mu}$ ($a_{\mu}$) is the creation (annihilation) operator
of electron in state labelled by ``$\mu$",
which labels both the multi-orbital
and distinct spin states of the system.
The second term is the Hamiltonian of the two electrodes, which
are also termed as emitter (left electrode) and collector (right
electrode), in some places of this work as usual. The third term
describes tunneling between the electrodes and the system.
In this paper the electrode reservoir electrons are also
attached with the index ``$\mu$" to characterize their possible
correspondence to the {\it system states}. For instance, this
will be the typical situation in the spin-dependent transport.

\begin{figure}\label{Fig1}
\begin{center}
\centerline{\includegraphics [scale=0.25]{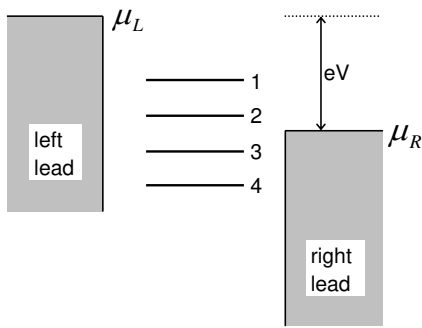}}
\caption{ Schematic setup for electrical transport
through a multi-level mesoscopic system. }
\end{center}
\end{figure}

Introducing the reservoir operators
$F_{\mu} = \sum_{\al}\sum_{k} t_{\al\mu k}d_{\al\mu k}
  \equiv f_{L\mu} + f_{R\mu}$,
we re-express the tunneling Hamiltonian as
\begin{eqnarray}\label{Ht}
H' &=& \sum_{\mu} \left( a^{\dg}_{\mu} F_{\mu}
       + \rm{H.c.}\right) .
\end{eqnarray}
Then, regarding this tunneling Hamiltonian as perturbation,
the second-order cummulant expansion leads us to
a formal equation for the reduced density matrix \cite{Yan98}
\bea\label{ME-1}
\dot{\rho}(t)
= -i {\cal L}\rho(t) - \int^{t}_{0}d\tau \la {\cal L}'(t){\cal G}(t,\tau)
                {\cal L}'(\tau){\cal G}^{\dg}(t,\tau) \ra \rho(t).
\eea
Here the Liouvillian superoperators are defined as
${\cal L}(\cdots)\equiv [H_S,(\cdots)]$,
${\cal L'}(\cdots)\equiv [H',(\cdots)]$, and
${\cal G}(t,\tau)(\cdots)\equiv G(t,\tau)(\cdots)G^{\dg}(t,\tau)$
with $G(t,\tau)$ the usual propagator (Green's function) associated with
the {\it system} Hamiltonian $H_S$.
The reduced density matrix $\rho(t)=\rm{Tr}_B[\rho_T(t)]$,
and $\la (\cdots)\ra=\rm{Tr}_B[(\cdots)\rho_B]$ with $\rho_B$
the density matrix of the electron reservoirs.
Notice that \Eq{ME-1} is nothing but an alternative form of the
quantum master equation under the second-order Born approximation.
The underlining assumption is that the tunneling Hamiltonian is
not strong enough, which makes the second-order cummulant expansion
reasonable. It is known that this approximation
applies well to most dissipative systems in quantum optics.
Noticeably, for most transport systems, weak tunneling is also the
typical regime where various forms of golden-rule type theories
are adopted.
In strong tunneling regime, special technique is required,
which goes beyond the present second-order Born approximation,
and other golden-rule type theories \cite{Sch94}.

The trace in \Eq{ME-1} is over all the electrode degrees of freedom,
leading thus to the equation of motion of the {\it unconditional}
reduced density matrix of the system.
To describe the transport problem, we should keep track of the record
of electron numbers arrived at the collector (emitted from the emitter
and passed through the mesoscopic system
in between the two electrodes).
We therefore classify the Hilbert space of the electrodes as follows.
First, we define the subspace in the absence of electron
arrived at the collector as
``$B^{(0)}$", which is spanned by the product
of all many-particle states of the two isolated reservoirs, formally denoted
as $B^{(0)}\equiv\mb{span}\{|\Psi_L\ra\otimes |\Psi_R\ra \}$.
Then, we introduce the Hilbert subspace ``$B^{(n)}$" ( $n=1,2,\cdots$),
corresponding to $n$-electrons arrived at the collector.
The entire Hilbert space of the two electrodes is $B=\oplus_n B^{(n)}$.

With the above classification of the reservoir states, the average over states
in the entire Hilbert space ``$B$" in \Eq{ME-1}
is replaced with states in the subspace ``$B^{(n)}$",
leading to a {\it conditional} master equation
\bea\label{ME-2}
\dot{\rho}^{(n)}(t) &=& -i{\cal L}\rho^{(n)}(t) - \int^{t}_{0}d\tau
      \mb{Tr}_{B^{(n)}} [{\cal L}'(t){\cal G}(t,\tau)   \nl
      & &  \times {\cal L}'(\tau) {\cal G}^{\dg}(t,\tau) \rho_T(t)] .
\eea
Here $\rho^{(n)}(t)=\mb{Tr}_{B^{(n)}}[\rho_T(t)]$,
which is the reduced density matrix of the system
{\it conditioned} by the number of electrons arrived at the collector
until time $t$.
Now we transform the Liouvillian operator product in \Eq{ME-2}
into the conventional Hilbert form:
\bea\label{L-H}
&&  {\cal L}'(t){\cal
G}(t,\tau) {\cal L}'(\tau) {\cal G}^{\dg}(t,\tau) \rho_T(t) \nl
&=& [ H'(t)G(t,\tau)H'(\tau)G^{\dg}(t,\tau)\rho_T(t)  \nl & &
- G(t,\tau)H'(\tau)G^{\dg}(t,\tau)\rho_T(t)H'(t)] + \mb{H.c.} \nl
&\equiv& [I-II]+\mb{H.c.}
\eea
To proceed, two physical considerations are further implemented as follows:
(i)
Instead of the conventional Born approximation for the entire density matrix
$\rho_T(t)\simeq\rho(t)\otimes\rho_B$,
we propose the ansatz $\rho_T(t)\simeq\sum_n\rho^{(n)}(t)\otimes\rho_B^{(n)}$,
where $\rho_B^{(n)}$ is the density operator of the electron reservoirs
associated with $n$-electrons arrived at the collector.
With this ansatz for the density operator, tracing over the subspace
``$B^{(n)}$" yields
\begin{subequations}\label{n-ave}
\bea \mb{Tr}_{B^{(n)}}[I]&=& \sum_{\mu,\nu} \left\{
           \mb{Tr}_B [F_{\mu}^{\dg}(t)F_{\nu}(\tau)\rho_B^{(n)}] \right. \nl
       & & \times [a_{\mu}G(t,\tau)a_{\nu}^{\dg}G^{\dg}(t,\tau)\rho^{(n)}]
\nl
       & &  +   \mb{Tr}_B [F_{\mu}(t)F_{\nu}^{\dg}(\tau)\rho_B^{(n)}]  \nl
       & & \left. \times [a_{\mu}^{\dg}G(t,\tau)a_{\nu}G^{\dg}(t,\tau)\rho^{(n)}]
           \right\}  \\
\mb{Tr}_{B^{(n)}}[II]&=& \sum_{\mu,\nu} \left\{
           \mb{Tr}_B[f_{L\nu}^{\dg}(\tau)\rho_B^{(n)}f_{L\mu}(t)] \right. \nl
       & & \times [ G(t,\tau)a_{\nu}G^{\dg}(t,\tau)\rho^{(n)}a_{\mu}^{\dg} ]   \nl
       & & + \mb{Tr}_B [f_{L\nu}(\tau)\rho_B^{(n)}f_{L\mu}^{\dg}(t)] \nl
       & & \times [ G(t,\tau)a_{\nu}^{\dg}G^{\dg}(t,\tau)\rho^{(n)}a_{\mu} ] \nl
       &  &  +\mb{Tr}_B[f_{R\nu}^{\dg}(\tau)\rho_B^{(n-1)}f_{R\mu}(t)] \nl
       & & \times [ G(t,\tau)a_{\nu}G^{\dg}(t,\tau)\rho^{(n-1)}a_{\mu}^{\dg} ]   \nl
       & & + \mb{Tr}_B [f_{R\nu}(\tau)\rho_B^{(n+1)}f_{R\mu}^{\dg}(t)] \nl
       & & \left. \times [ G(t,\tau)a_{\nu}^{\dg}G^{\dg}(t,\tau)
            \rho^{(n+1)}a_{\mu} ] \right\}.
\eea
\end{subequations}
Here we have utilized the orthogonality between states in different subspaces,
which in fact leads to the term selection from the entire density operator $\rho_T$.
(ii) Due to the closed nature of the transport circuit, the extra
electrons arrived at the collector (right reservoir) will flow back into
the emitter (left reservoir) via the external circuit. Also, the rapid
relaxation processes in the reservoirs will quickly bring the
reservoirs to the local thermal equilibrium state determined by
the chemical potentials. As a consequence, after the procedure
(i.e. the state selection) done in \Eq{n-ave}, the electron reservoir
density matrices $\rho_B^{(n)}$ and $\rho_B^{(n\pm 1)}$
should be replaced by $\rho_B^{(0)}$, i.e., the
local thermal equilibrium reservoir state,
which leads the reservoir correlation functions in \Eq{n-ave} to be,
respectively,
$\la f_{\alpha\mu}^{\dg}(t)f_{\alpha\nu}(\tau)\ra =
C_{\alpha\mu\nu}^{(+)}(t-\tau)$, and $\la
f_{\alpha\mu}(t)f^{\dg}_{\alpha\nu}(\tau)\ra =
C_{\alpha\mu\nu}^{(-)}(t-\tau)$.
Here $\la \cdots \ra$ stands for $\mb{Tr}_B [(\cdots)\rho_B^{(0)}]$,
with the usual meaning of thermal average.
Obviously, $\la F_{\mu}^{\dg}(t)F_{\nu}(\tau)\ra
= C_{\mu\nu}^{(+)}(t-\tau)=\sum_{\alpha=L,R}C_{\alpha\mu\nu}^{(+)}(t-\tau)$,
and $\la F_{\mu}(t)F^{\dg}_{\nu}(\tau)\ra = C_{\mu\nu}^{(-)}(t-\tau)
=\sum_{\alpha=L,R}C_{\alpha\mu\nu}^{(-)}(t-\tau)$.

Under the Markovian approximation, the time integral in \Eq{ME-2}
is replaced by $\frac{1}{2}\int^{\infty}_{-\infty}d\tau$.
This approximation considerably simplifies the result.
For instance, substituting the first term of \Eq{n-ave}
into the time integral of \Eq{ME-2},
we have
$\int^{\infty}_{-\infty}d\tau C_{\mu\nu}^{(+)}(t-\tau)a_{\mu}
 [e^{-i{\cal L}(t-\tau)}a^{\dg}_{\nu}]\rho^{(n)}(t)
=a_{\mu}[C_{\mu\nu}^{(+)}(-{\cal L})a^{\dg}_{\nu}]\rho^{(n)}(t)$.
Other terms can be similarly integrated out, leading to
\bea\label{ME-3}
\dot{\rho}^{(n)}
   &=&  -i {\cal L}\rho^{(n)} - \frac{1}{2} \sum_{\mu}
   \left\{ [a_{\mu}^{\dg}A_{\mu}^{(-)}\rho^{(n)}
   +\rho^{(n)}A_{\mu}^{(+)}a_{\mu}^{\dg}
\right. \nl & &         - A_{L\mu}^{(-)}\rho^{(n)}a_{\mu}^{\dg}
            - a_{\mu}^{\dg}\rho^{(n)}A_{L\mu}^{(+)}        \nl
& & \left. - A_{R\mu}^{(-)}\rho^{(n-1)}a_{\mu}^{\dg}
   - a_{\mu}^{\dg}\rho^{(n+1)}A_{R\mu}^{(+)}]+{\rm H.c.} \right\} .
\eea
Here $A_{\alpha\mu}^{(\pm)}=\sum_{\nu}
C_{\alpha\mu\nu}^{(\pm)}(\pm {\cal L})a_{\nu}$,
and $A_{\mu}^{(\pm)}=\sum_{\alpha=L,R}A_{\alpha\mu}^{(\pm)}$.
The spectral functions $C_{\alpha\mu\nu}^{(\pm)}(\pm {\cal L})$
are defined in terms of the Fourier transform of the reservoir
correlation functions, i.e.,
$ C_{\alpha\mu\nu}^{(\pm)}(\pm {\cal L})=\int^{\infty}_{-\infty} dt
  C_{\alpha\mu\nu}^{(\pm)}(t) e^{\pm i{\cal L}t}$.
We would like to remark here
that this time integral leads to ``exact" energy conservation
for electron transfer between the electrodes and the central system.
This ``conservation law" would cause errors in the near-resonance bias.
For instance, at zero temperature and for electrode Fermi level(s)
lower than but {\it very close} to certain system level,
the present ``energy conservation law" does not permit any electron
occupation on the concerned system level.
Nevertheless, the nGF-based quantum kinetic theory allows
occupation under the same condition \cite{Dav93}.
The underlying reason is the neglect of level broadening
in present treatment, whose inclusion is referred to the Appendix.

The ``$n$"-dependence of \Eq{ME-3} is analogical to the usual
rate equation, despite its formal matrix/operator feature.
Each term of \Eq{ME-3} can be similarly interpreted as
for the conventional ``c-number" rate equation.
Compared with the Bloch equation derived by Gurvitz
{\it et al} \cite{Gur96b}, in \Eq{ME-3} $\rho^{(n)}$ is also
coupled to $\rho^{(n+1)}$ which is absence from
Ref.\ \onlinecite{Gur96b}. This feature originates from the general
nature that \Eq{ME-3} is established under
non-zero temperature and arbitrary (not necessarily large) bias voltage.

With the knowledge of $\rho^{(n)}(t)$, one is readily able to compute
the various transport properties, such as the transport current and
noise spectrum \cite{Li04b}.
Remarkably, we can derive a compact expression for the current
which is only related to the unconditional
density matrix $\rho(t)=\sum_n\rho^{(n)}(t)$.
The derivation is started with the physical observation that
the current can be determined by the probability distribution
function $P(n,t)\equiv\rm{Tr}[\rho^{(n)}(t)]$, in terms of
$I(t)=ed \bar{N}(t) /dt$, where $\bar{N}(t)=\sum_n nP(n,t)$.
Properly treating the summation over ``$n$" and making use of the
cyclic property under trace,
minor algebra based on \Eq{ME-3} straightforwardly leads to
\bea\label{I-t}
I(t) &=& e \sum_n n \rm{Tr} \left[ \dot{\rho}^{(n)}(t)\right] \nl
     &=& \frac{e}{2} \sum_{\mu} \rm{Tr} \left[ \left(
         a^{\dg}_{\mu}A^{(-)}_{R\mu}-A^{(+)}_{R\mu}a^{\dg}_{\mu}
         \right)\rho(t)+\rm{H.c.} \right] .
\eea
Here the unconditional density matrix $\rho=\sum_n \rho^{(n)}$
satisfies an even simpler equation, which can be easily derived by
summing up \Eq{ME-3} over ``$n$"
\bea\label{rho-t}
 \dot{\rho} = -i {\cal L}\rho - \frac{1}{2}\sum_{\mu}
        \left\{ [a_{\mu}^{\dg},A_{\mu}^{(-)}\rho
        -\rho A_{\mu}^{(+)}]+ \rm{H.c.} \right\}  .
\eea
\Eqs{I-t} and (\ref{rho-t}) together with (\ref{ME-3})
constitute the principal result of this work,
which can serve as a convenient starting point
to compute transport current under wide range of conditions,
such as in the presence of many-body
Coulomb interaction, at finite temperatures and for arbitrary voltages.
Moreover, the current expression and the associated master equation
are free from state representation and the specific system Hamiltonian,
which therefore holds the merit of unification in its applications.
For instance, for quantum transport through an interacting system,
which is usually a challenging problem, one can first diagonalize
the {\it isolated} system Hamiltonian, then do the Liouvillian
operation easily in the eigen-state representation.
In the following, as application of this approach
we only illustrate a number of simple examples,
and remain the systematic applications to more interesting
problems to be the subject of forthcoming works.

\section{Illustrative Applications}

\subsection{Single Level System}

As a preliminary application of \Eq{ME-3}, let us consider the
resonant transport through a single level system.
Under wide-band approximation for the electrodes,
the reservoir electron correlation functions read
$ C_{\alpha}^{(\pm)}(t-\tau)=|t_{\alpha}|^2\sum_k
  e^{\pm i\ep_k(t-\tau)}
               n^{(\pm)}_{\alpha}(\ep_k) $,
where $n^{(+)}_{\alpha}(\ep_k)= n_{\alpha}(\ep_k)$
is the Fermi distribution function, and
$n^{(-)}_{\alpha}(\ep_k)= 1-n_{\alpha}(\ep_k)$.
Then the spectral function can be easily carried out as
\bea\label{C+-L}
A_{\al}^{(\pm)}=C_{\alpha}^{(\pm)}(\pm{\cal L})a
=\Gamma_{\alpha}n^{(\pm)}_{\alpha}(E_0)a .
\eea
Here, $\Gamma_{\alpha}=2\pi g_{\alpha}|t_{\alpha}|^2$,
with $g_{\alpha}$ the density of states of the ``$\alpha$" electrode.
In the special case of zero temperature and large bias voltage
$\mu_L\gg E_0\gg \mu_R$, which is in fact the {\it applicable condition}
of Gurvitz's approach \cite{Gur96b}, we simply have
$A^{(+)}_{L} = \Gamma_L a$, $A^{(-)}_{L} = 0$,
$A^{(-)}_{R} = \Gamma_R a$, and $A^{(+)}_{R} = 0$.
Substituting these into \Eq{ME-3} yields
\bea\label{ME-RT}
\dot{\rho}^{(n)}
   &=&  -i {\cal L}\rho^{(n)} - \frac{1}{2}
   \left\{ [\Gamma_R a^{\dg}a\rho^{(n)}+\Gamma_L\rho^{(n)}aa^{\dg} \right. \nl
   & & \left. - \Gamma_L a^{\dg}\rho^{(n)}a
   -  \Gamma_R a\rho^{(n-1)}a^{\dg}]+{\rm H.c.} \right\} .
\eea
To obtain the matrix element form of this equation,
let us choose the empty (level) state $|0\ra$ and the occupied state
$|1\ra$ as representation basis.
Straightforwardly, by computing the matrix elements of the terms of
\Eq{ME-RT} one by one, we obtain
\bea\label{rho-0011}
\dot{\rho}^{(n)}_{00}&=&-\Gamma_L\rho^{(n)}_{00}+\Gamma_R\rho^{(n-1)}_{11}, \nl
\dot{\rho}^{(n)}_{11}&=&-\Gamma_R\rho^{(n)}_{11}+\Gamma_L\rho^{(n)}_{00}.
\eea
This is the result derived by Gurvitz {\it et al}
under the limits mentioned above \cite{Gur96b}.


\subsection{Multi-Level System}

Now we consider the transport through a multi-level system as shown in
Fig.\ 1, under arbitrary voltage and at finite temperature.
The system Hamiltonian simply reads
$H_S=\sum^{N}_{\mu=1}E_{\mu}a_{\mu}^{\dg}a_{\mu}$.
Also, let us assume that the level separation
is much larger than the characteristic level widths, i.e.,
$|E_{\mu}-E_{\mu-1}|\gg\Gamma_L,\Gamma_R$,
which leads to the correlation function
$C^{(\pm)}_{\alpha\mu\nu}(t)\simeq\delta_{\mu\nu}C^{(\pm)}_{\alpha\mu\mu}(t)$.
This assumption neglects the interference effect of electron tunneling
through different levels, which is significant only in the case
$|E_{\mu}-E_{\mu-1}|<\Gamma_L,\Gamma_R$.
Similar to single level system, we have
$A_{R\mu}^{(\pm)}=\Gamma_R(E_{\mu})n^{(\pm)}_R(E_{\mu})a_{\mu}$.
For this simplified model, the reduced system density matrix
is the direct product of the individual single level density matrix,
i.e., $\rho=\otimes^{N}_{\mu=1}\rho_{\mu}$,
and the steady-state solution
of the single level density matrix can be easily obtained as
$\rho_{\mu}=p_{\mu}|1\ra_{\mu} \la 1|+(1-p_{\mu})|0\ra_{\mu} \la 0|$,
where $|1\ra_{\mu}$ ($|0\ra_{\mu}$) stands for the occupied
(unoccupied) state of the $\mu$th level,
and the occupation probability $p_{\mu}$ reads
\bea
p_{\mu}=\frac{n_{L}(E_{\mu})\Gamma_{L}(E_{\mu})
        +n_{R}(E_{\mu})\Gamma_{R}(E_{\mu})}
        {\Gamma_{L}(E_{\mu})+\Gamma_{R}(E_{\mu})}.
\eea
Substituting the obtained $A_{R\mu}^{(\pm)}$ and $\rho$
into the current expression \Eq{I-t}, we arrive at an expression
for the steady-state current as
\bea\label{IV-1}
I = e \sum_{\mu}\frac{\Gamma_{L}(E_{\mu})\Gamma_{R}(E_{\mu})}
        {\Gamma_{L}(E_{\mu})+\Gamma_{R}(E_{\mu})}
        \left[n_{L}(E_{\mu})-n_{R}(E_{\mu})  \right] .
\eea
This result clearly manifests the typical step-like I-V characteristics,
where each step corresponds to involving
a new level into the conduction by increasing the bias voltage ,
with the standard resonant current
$e\Gamma_L\Gamma_R/(\Gamma_L+\Gamma_R)$.

\subsection{Non-interacting Coupled Quantum Dots}

In the above multi-level system the quantum coherence or nature of
quantum superposition of system states is not manifested,
and the result can be obtained via {\it classical} rate equations.
To reveal more clearly the {\it quantum} nature of the developed formalism,
in this subsection we consider transport through system
of a coupled quantum dots \cite{Gur96b}.
In this case, the non-diagonal elements of density matrix,
which have no classical counterparts, will appear in the equations
of motion and play essential role.

The Hamiltonian of the coupled quantum dots reads
$H_S=E_1a^{\dg}_1a_1+E_2a^{\dg}_2a_2
  +\Omega(a^{\dg}_1a_2+a^{\dg}_2a_1)$,
where each dot contains a single resonant level $E_1$ ($E_2$),
and the two dots are coupled by $\Omega$.
In principle, for any system the master equation
(\ref{ME-3}) or (\ref{rho-t}) can be expressed and solved
in the system eigenstate representation.
Here, for the coupled quantum dots, we would like to present a more
elegant method in terms of the language of Bogoliubov transformation,
to explicitly carry out the superoperators $A^{(\pm)}_{\alpha\mu}$.
To diagonalize $H_S$, the standard Bogoliubov transformation
defines a pair of new electron operators as follows:
$b_1=ua_1+va_2$, and $b_2=ua_2-va_1$.
The desired diagonalized Hamiltonian reads
$H_S=\ti{E}_1b^{\dg}_1b_1+\ti{E}_2b^{\dg}_2b_2$.
The diagonalization condition $(E_2-E_1)uv+\Omega(u^2-v^2)=0$,
together with the normalization condition $u^2+v^2=1$,
uniquely determine the transformation coefficients $u$ and $v$,
and the eigen-energies read accordingly,
$\ti{E}_1=E_1u^2+E_2v^2+2\Omega uv$, and
$\ti{E}_2=E_1v^2+E_2u^2-2\Omega uv$.
Simple algebra leads to
${\cal L}^na_1=(-\ti{E}_1)^n ub_1-(-\ti{E}_2)^n vb_2$,
and ${\cal L}^na_2=(-\ti{E}_1)^n vb_1+(-\ti{E}_2)^n ub_2$.
Notice that in the wide-band approximation for the electrode reservoirs,
$C^{(\pm)}_{\alpha}(\pm {\cal L})=\Gamma_{\alpha}
 n^{(\pm)}_{\alpha}(-{\cal L})$. We thus have
\bea
A^{(\pm)}_{L} &=& \Gamma_L\left[u n^{(\pm)}_L(\ti{E}_1)b_1
                           -v n^{(\pm)}_L(\ti{E}_2)b_2\right] ,\nl
A^{(\pm)}_{R} &=& \Gamma_R\left[v n^{(\pm)}_R(\ti{E}_1)b_1
                           +u n^{(\pm)}_R(\ti{E}_2)b_2\right] .
\eea
With this result, the explicit form of the master equation can be
easily obtained for arbitrary offset of the dot levels ($E_1$ and $E_2$).
To compare with the Bloch equations derived by Gurvitz {\it et al}
\cite{Gur96b}, consider the special configuration of the
two dot levels in resonance, i.e., $E_1=E_2\equiv E_0$.
For this setup,
$u=-v=1/\sqrt{2}$, and $\ti{E}_{1,2}=E_0\mp \Omega$.
Moreover,
in the large bias voltage limit $\mu_L\gg \ti{E}_2,\ti{E}_1\gg\mu_R$,
we simply have
$A^{(+)}_{L}=\Gamma_L a_1$, $A^{(-)}_{L}=0$,
$A^{(+)}_{R}=0$, and $A^{(-)}_{R}=\Gamma_R a_2$.
Substituting them into \Eq{ME-3}, an explicit form of conditional
master equation is obtained as
\bea\label{ME-2dots}
\dot{\rho}^{(n)}
    &=&  -i {\cal L}\rho^{(n)} - \frac{1}{2}
         \left\{ [\rho^{(n)}\Gamma_L a_1a_1^{\dg}
         + a_2^{\dg}\Gamma_R a_2 \rho^{(n)} \right.  \nl
    & & \left. -a_1^{\dg}\rho^{(n)}\Gamma_L a_1
        - \Gamma_R a_2 \rho^{(n-1)} a_2^{\dg} ]
        +{\rm H.c.}\right\} .
\eea
In the electron number representation
$\{ |1\ra, |2\ra, |3\ra, |4\ra\}$,
which correspond to, respectively, the states of
no electron in the two dots, one electron in the left (right) dot,
and two electrons in each dot,
\Eq{ME-2dots} can be precisely recast to the result derived
in Ref.\ \onlinecite{Gur96b}, where the quantum coherence nature
beyond the classical rate equation was particularly emphasized.


\subsection{Single Level System in the Presence of Charging Effect}

The above examples do not involve many-electron Coulomb interaction.
In this subsection, we consider the simplest example of transport
through single level system in the presence of Coulomb charging effect.
The system Hamiltonian reads
$H_S=\sum_{\mu}(E_0+\frac{U}{2}n_{\bar{\mu}})n_{\mu}$.
Here the index $\mu$ labels the spin up (``$\uparrow$") and
spin down (``$\downarrow$") states,
and $\bar{\mu}$ stands for the opposite spin orientation.
The electron number operator $n_{\mu}=a^{\dg}_{\mu}a_{\mu}$,
and the Hubbard term $Un_{\uparrow}n_{\downarrow}$ describe
the charging effect.
Obviously, the reservoir correlation
function is diagonal with respect to the spin indices, i.e.,
$C^{(\pm)}_{\alpha\mu\nu}(t)=\delta_{\mu\nu}C^{(\pm)}_{\alpha\mu\mu}(t)$.
We thus have
\bea\label{A+-CBE}
A^{(\pm)}_{\alpha\mu}
  = C^{(\pm)}_{\alpha\mu\mu}(\pm {\cal L})a_{\mu}
  = C^{(\pm)}_{\alpha\mu\mu}[\mp(E_0+Un_{\bar{\mu}})]a_{\mu}  .
\eea
Moreover, for either spin-up or spin-down electrons,
the spectral functions $C^{(\pm)}_{\alpha\mu\mu}(\pm {E})$
are identical to \Eq{C+-L}.
For present system, the four basis states can be chosen as
$|1\ra=|00\ra_{\uparrow\downarrow}$,
$|2\ra=|10\ra_{\uparrow\downarrow}$,
$|3\ra=|01\ra_{\uparrow\downarrow}$, and
$|4\ra=|11\ra_{\uparrow\downarrow}$.
Also in the limiting case of zero temperature and large bias voltage
($\mu_L\gg E_0+U>E_0\gg\mu_R$),
inserting \Eq{A+-CBE} into \Eq{ME-3} and carrying out the matrix elements
associated with the above four basis states, we obtain
\begin{subequations}\label{n-rho-dot}
\bea \dot{\rho}_{11}^{(n)}
   &=&  -2\Gamma_L\rho_{11}^{(n)}+\Gamma_R\rho_{22}^{(n-1)}
        + \Gamma_R\rho_{33}^{(n-1)}  ,  \\
\dot{\rho}_{22}^{(n)}
   &=&  -(\Gamma_R+\Gamma'_L)\rho_{22}^{(n)}+\Gamma_L\rho_{11}^{(n)}
        + \Gamma'_R\rho_{44}^{(n-1)}  ,  \\
\dot{\rho}_{33}^{(n)}
   &=&  -(\Gamma_R+\Gamma'_L)\rho_{33}^{(n)}+\Gamma_L\rho_{11}^{(n)}
        + \Gamma'_R\rho_{44}^{(n-1)}  ,  \\
\dot{\rho}_{44}^{(n)}
   &=&  -2\Gamma'_R\rho_{44}^{(n)}+\Gamma'_L\rho_{22}^{(n)}
        + \Gamma'_L\rho_{33}^{(n)}  ,
\eea
\end{subequations}
where $\Gamma'_{\alpha}=(2\pi g_{\alpha}|t_{\alpha}|^2)_{\tiny{E=E_0+U}}$.
Satisfactorily, \Eq{n-rho-dot} is nothing but the result obtained
in Ref.\ \onlinecite{Gur96b} under the same limiting conditions.

\subsection{Interacting Quantum Dot with Zeeman Splitting}

In this subsection, we revisit the model studied in Sec. III-D,
but slightly modify it by allowing for a finite spin splitting, i.e.,
$H_S=\sum_{\mu=\uparrow,\downarrow}(E_{\mu}a_{\mu}^{\dg}a_{\mu}
 +\frac{U}{2}n_{\mu}n_{\bar{\mu}})$,
where the non-zero Zeeman splitting is characterized by
$E_{\downarrow}-E_{\uparrow}\equiv\Delta$.
The transport properties of
this system has been studied recently by Thielmann {\it et al} \cite{Thi},
by applying the real-time diagrammatic technique \cite{Sch94}.
Here we show our master equation approach
can solve this non-trivial model in a more transparent way.

As done previously, we first carry out the commutator
${\cal L}a_{\mu}\equiv[H_S,a_{\mu}]=-W_{\mu}a_{\mu}$,
where $W_{\mu}=E_{\uparrow}\delta_{\uparrow\mu}
       +E_{\downarrow}\delta_{\downarrow\mu}
       +U(n_{\uparrow}\delta_{\downarrow\mu}
          +n_{\downarrow}\delta_{\uparrow\mu})$.
Noting that $[H_S,W_{\mu}]=0$,
we have ${\cal L}^na_{\mu}=(-W_{\mu})^na_{\mu}$.
Accordingly, $A^{(\pm)}_{\alpha\mu}=C^{(\pm)}_{\alpha\mu\mu}
(\mp W_{\mu})a_{\mu}$.
In the wide-band approximation and assuming an energy-independent
coupling strength $\Gamma_L$ ($\Gamma_R$) with the left (right) electrode,
explicit expressions for $A^{(\pm)}_{\alpha\mu}$ are obtained as
$A^{(\pm)}_{L/R\mu}=\Gamma_{L/R} n^{(\pm)}_{L/R}(W_{\mu})a_{\mu}$.
In the occupation number representation, i.e.,
$|1\ra=|00\ra_{\uparrow\downarrow}$,
$|2\ra=|10\ra_{\uparrow\downarrow}$,
$|3\ra=|01\ra_{\uparrow\downarrow}$, and
$|4\ra=|11\ra_{\uparrow\downarrow}$,
the master equation \Eq{rho-t} can be easily solved,
and via \Eq{I-t} the current can be computed quite straightforwardly.
In the following, we explicitly carry out the result
in different voltage regimes.
For the sake of being able to obtain analytic result, we focus on
the limiting case of zero temperature.
Moreover, without loss of generality, we assume
that the bias voltage makes
the Fermi level of the right electrode be always lower
than the quantum dot energy levels during transport.
Therefore, all $n^{(+)}_R$ at the four energies,
say, $E_{\uparrow}$, $E_{\downarrow}$, $E_{\uparrow}+U$,
and $E_{\downarrow}+U$, are zero.

{\it Regime (i)}:
$\mu_L>E_{\uparrow}+U,E_{\downarrow}+U,E_{\uparrow},E_{\downarrow}>\mu_R$.
In this high bias regime, the corresponding Fermi functions are
$n^{(+)}_L(E_{\uparrow})=n^{(+)}_L(E_{\downarrow})
=n^{(+)}_L(E_{\uparrow}+U)=n^{(+)}_L(E_{\downarrow}+U)=1$, and
the master equation \Eq{rho-t} reads
\bea \dot{\rho}_{11}
   &=&  -2\Gamma_L\rho_{11}+\Gamma_R\rho_{22}
        + \Gamma_R\rho_{33}  ,  \nl
\dot{\rho}_{22}
   &=&  -(\Gamma_R+\Gamma_L)\rho_{22}+\Gamma_L\rho_{11}
        + \Gamma_R\rho_{44}  ,  \nl
\dot{\rho}_{33}
   &=&  -(\Gamma_R+\Gamma_L)\rho_{33}+\Gamma_L\rho_{11}
        + \Gamma_R\rho_{44}  ,  \nl
\dot{\rho}_{44}
   &=&  -2\Gamma_R\rho_{44}+\Gamma_L\rho_{22}
        + \Gamma_L\rho_{33}  .
\eea
To evaluate the stationary current, only stationary solution
is required, which are easily obtained as, respectively,
$\rho_{22}=\rho_{33}=\Gamma_L\Gamma_R/(\Gamma_L+\Gamma_R)^2$,
$\rho_{11}=(\Gamma_R/\Gamma_L)\rho_{22}$, and
$\rho_{44}=(\Gamma_L/\Gamma_R)\rho_{22}$.
Then, from \Eq{I-t} the current is simply carried out as
\bea
I(t\rightarrow\infty)
  = e\Gamma_R(\rho_{22}+\rho_{33}+2\rho_{44})
  = \frac{2e\Gamma_L\Gamma_R}{\Gamma_L+\Gamma_R} .
\eea

{\it Regime (ii)}:
$E_{\uparrow}+U>\mu_L>E_{\downarrow}+U,E_{\uparrow},E_{\downarrow}>\mu_R$.
The Fermi functions in this case read
$n^{(+)}_L(E_{\uparrow}+U)=0$,
$n^{(+)}_L(E_{\uparrow})=n^{(+)}_L(E_{\downarrow})
=n^{(+)}_L(E_{\downarrow}+U)=1$, and
the resulting master equation is
\bea
\dot{\rho}_{11}
   &=&  -2\Gamma_L\rho_{11}+\Gamma_R\rho_{22}
        + \Gamma_R\rho_{33}  ,  \nl
\dot{\rho}_{22}
   &=&  -(\Gamma_R+\Gamma_L)\rho_{22}+\Gamma_L\rho_{11}
        + \Gamma_R\rho_{44}  ,  \nl
\dot{\rho}_{33}
   &=&  -\Gamma_R\rho_{33}+\Gamma_L\rho_{11}
        + \Gamma_R\rho_{44}+\Gamma_L\rho_{44}  ,  \nl
\dot{\rho}_{44}
   &=&  -(\Gamma_L+2\Gamma_R)\rho_{44}+\Gamma_L\rho_{22} .
\eea
Solution of the stationary state reads, respectively,
$\rho_{22}=\Gamma_L\Gamma_R(\Gamma_L+2\Gamma_R)/2(\Gamma_L+\Gamma_R)^3$,
$\rho_{33}=\rho_{11}+2\Gamma_L^2(\Gamma_L+\Gamma_R)/2(\Gamma_L+\Gamma_R)^3$,
and $\rho_{44}=\Gamma_L^2\Gamma_R/2(\Gamma_L+\Gamma_R)^3$.
Note that $\rho_{11}=1-\rho_{22}-\rho_{33}-\rho_{44}$, which
is irrelevant to the current.
Straightforwardly, we obtain the current
\bea
I(t\rightarrow\infty)
  &=& e\Gamma_R(\rho_{22}+\rho_{33}+2\rho_{44}) \nl
  &=& \frac{e\Gamma_L\Gamma_R(\Gamma_L+2\Gamma_R)}{(\Gamma_L+\Gamma_R)^2} .
\eea

{\it Regime (iii)}:
$E_{\uparrow}+U,E_{\downarrow}+U >\mu_L>E_{\uparrow},E_{\downarrow}>\mu_R$.
The Fermi functions $n^{(+)}_L(E_{\downarrow})=n^{(+)}_L(E_{\uparrow})=1$,
and $n^{(+)}_L(E_{\uparrow}+U)=n^{(+)}_L(E_{\downarrow}+U)=0$.
The corresponding master equation reads
\bea
\dot{\rho}_{11}
   &=&  -2\Gamma_L\rho_{11}+\Gamma_R\rho_{22}
        + \Gamma_R\rho_{33}  ,  \nl
\dot{\rho}_{22}
   &=&  -\Gamma_R\rho_{22}+\Gamma_L\rho_{11}
        + (\Gamma_L+\Gamma_R)\rho_{44}  ,  \nl
\dot{\rho}_{33}
   &=&  -\Gamma_R\rho_{33}+\Gamma_L\rho_{11}
        + (\Gamma_L+\Gamma_R)\rho_{44}  ,  \nl
\dot{\rho}_{44}
   &=&  -2(\Gamma_L+\Gamma_R)\rho_{44} .
\eea
The stationary-state solution of the reduced density matrix
leads to the transport current as
\bea
I(t\rightarrow\infty)
  = e\Gamma_R(\rho_{22}+\rho_{33}+\rho_{44})
  = \frac{2e\Gamma_L\Gamma_R}{2\Gamma_L+\Gamma_R} .
\eea

{\it Regime (iv)}:
$E_{\uparrow}+U,E_{\downarrow}+U,E_{\uparrow} >\mu_L>E_{\downarrow}>\mu_R$.
In this setup, only $n^{(+)}_L(E_{\downarrow})=1$,
and all other Fermi functions are zero.
Similarly, we first carry out the master equation
\bea
\dot{\rho}_{11}
   &=&  -\Gamma_L\rho_{11}+\Gamma_L\rho_{22}
        + \Gamma_R\rho_{22}+\Gamma_R\rho_{33}  ,  \nl
\dot{\rho}_{22}
   &=&  -(\Gamma_L+\Gamma_R)\rho_{22}
        + (\Gamma_L+\Gamma_R)\rho_{44}  ,  \nl
\dot{\rho}_{33}
   &=&  -\Gamma_R\rho_{33}+\Gamma_L\rho_{11}
        + (\Gamma_L+\Gamma_R)\rho_{44}  ,  \nl
\dot{\rho}_{44}
   &=&  -2(\Gamma_L+\Gamma_R)\rho_{44} .
\eea
Then, the stationary transport current is calculated via the
stationary-state solution of the density matrix as
\bea
I(t\rightarrow\infty)
  = e\Gamma_R(\rho_{22}+\rho_{33})
  = \frac{e\Gamma_L\Gamma_R}{\Gamma_L+\Gamma_R} .
\eea

Remarkably, we have precisely recovered all the Coulomb plateaus
presented in Ref.\ \onlinecite{Thi}, which go beyond simple intuition
and are obtained there by a not easily accessible real-time
diagrammatic technique.
This example may shine light on the convenience of our approach in applications.

\section{Concluding Remarks}

In summary, we have developed an efficient master equation approach
for quantum transport through mesoscopic systems,
and demonstrated its application by a number of examples.
Compared with the previous work by Gurvitz \cite{Gur96b}, the present
study not only generalizes the applicable conditions to
finite temperature and arbitrary voltage, but also identifies
the adopted approximation which appears not very clear in
Ref.\ \onlinecite{Gur96b}.
That is, by treating the electrodes as (Fermi) thermal baths,
the major approximation adopted in our derivation is
the standard second-order Born approximation for the coupling
(tunneling) Hamiltonian.
It is known that
this well-justified approximation makes the resultant quantum
master equation applicable in a large number of dissipative systems
(e.g. in quantum optics),
provided the system-bath coupling is not so strong.
Favorably, the illustrated examples in this paper also show
its applicability in quantum transport.
Moreover, the developed master equation approach holds the obvious advantages
of application convenience and straightforwardness, as well as
the ability to address many-body correlation, inelastic scattering,
and transient behavior, which are usually difficult issues
in mesoscopic transport.

In comparison with the nGF approach, we found that the structure of
\Eq{I-t} is in fact identical to the {\it formal expression}
of current in terms of non-equilibrium correlation functions \cite{Hau96}.
The nGF approach remains the relatively hard task to searching for particular
techniques (e.g. the Feynman diagram or equation of motion)
to carry out those correlation functions.
In this sense, the obtained \Eq{I-t} is nothing but the explicit
Markovian result under the second-order Born approximation
for the tunneling Hamiltonian. In principle,
further systematic corrections are possible along the line of going beyond
the Born approximation, to include higher order contribution of tunneling.
Finally, we mention that \Eq{I-t} can be derived from the
formal nGF expression of current, however, the present derivation
along the line of Ref.\ \onlinecite{Gur96b}
is interesting, and the particular result \Eq{ME-3}
from this unique method is of great value,
which contains rich information and
can be conveniently employed, for instance, to calculate the noise spectrum
\cite{Moz02,Li04b,Gur04}.

\appendix*
\section{Level-Broadening Effect}

To make the derived formulas \Eqs{ME-3}-(\ref{rho-t})
more accurately applicable to arbitrary voltage,
additional care is needed as any of the individual system levels
is approaching to the Fermi surface of the electrode.
For the sake of description clarity,
let us take the single-level resonance
system as an example to highlight the key point.
As mentioned previously,
the treatment in Sec.\ II under the second-order Born approximation
has neglected {\it level broadening} effect,
which would cause certain errors in some particular cases.
For instance, current flowing through the resonance system would
be strictly forbidden under the {\it near-resonance} condition,
i.e., as the resonance level $E_0$ is a little bit higher than
$\mu_L$ ($\mu_L>\mu_R$).
However, it is well known that a full quantum treatment will give
nonzero current in this situation \cite{Dav93,Hau96}.
For bias voltage such that the resonance level
(together with its broadening)
is within the range of the two Fermi levels (i.e. $\mu_L>E_0>\mu_R$),
common result of resonance current would be predicted by our master
equation approach and the nGF-based quantum transport theory.
In spite of this, it would be desirable to remove the drawback
of {\it inaccuracy} of our approach in the near-resonance situation.

To account for the level-broadening effect, we return to the evaluation
of $A^{(\pm)}_{\alpha\mu}=\sum_{\nu}
C^{(\pm)}_{\alpha\mu\nu}(\pm {\cal L})a_{\nu}$.
Without loss of generality, we restrict our description in the
{\it diagonal} case
$C^{(\pm)}_{\alpha\mu\nu}(t)=\delta_{\mu\nu}C^{(\pm)}_{\alpha\mu\mu}(t)$.
More general description is straightforward provided one has clarified
the correlation between ``$f_{\mu}$" and ``$f_{\nu}$".
Using the free-electron-gas model for the electrodes,
$A^{(\pm)}_{\alpha\mu}$ can be expressed as
\bea
A^{(\pm)}_{\alpha\mu}
  = 2\sum_k |t_{\alpha\mu,k}|^2 n_{\alpha}^{(\pm)}(\epsilon_k)
    \int^{\infty}_{0} dt e^{\pm i\epsilon_k t}
    e^{\pm i{\cal L}t} a_{\mu} .
\eea
In our previous treatment, we have replaced the time integral
$2\int_{0}^{\infty}dt$ by $\int_{-\infty}^{\infty}dt$,
under the spirit of Markovian approximation.
As a result, the time integration gives rise to
a $\delta$-function, $2\pi\delta(\epsilon_k+{\cal L})$,
which characterizes energy conservation for electron transfer
between the central system and the electrodes.
Mathematically, this procedure is equivalent to dropping the
imaginary (principal) part of integral, and keeping only the
real part.
Now, notice that $e^{\pm i{\cal L}t} a_{\mu}$ describes
the quantum evolution of the $E_{\mu}$ state (level) associated
with the {\it isolated} system Hamiltonian.
As a standard procedure,
the system level broadening effect due to coupling with the electrodes
can be implemented by inserting a damping factor
$e^{-\Gamma_{\mu}t}=e^{-(\Gamma_{L\mu}+\Gamma_{R\mu})t}$
into the integrand of the time integral \cite{note-1}.
After this, by keeping only the real part of the integral
and adopting the typical wide-band approximation
for the electrodes, we have
\bea\label{A+-2}
A^{(\pm)}_{\alpha\mu}
 = \Gamma_{\alpha\mu}\int\frac{d\epsilon_k}{2\pi}
   \ti{a}_{\alpha\mu}(\epsilon_k+{\cal L})
   n_{\alpha}^{(\pm)}(\epsilon_k)a_{\mu} .
\eea
Here the standard Lorentzian spectral density function reads
$\ti{a}_{\alpha\mu}(\omega+{\cal L})
 = 2\Gamma_{\mu}/[(\omega+{\cal L})^2+\Gamma^2_{\mu}]$.
Formally introducing
$N^{(\pm)}_{\alpha\mu}(-{\cal L})
\equiv \int\frac{\epsilon_k}{2\pi}
\ti{a}_{\alpha\mu}(\epsilon_k+{\cal L})
 n_{\alpha}^{(\pm)}(\epsilon_k)$,
we re-express (\ref{A+-2}) in a very compact form as
\bea\label{A+-3}
A^{(\pm)}_{\alpha\mu}
 = \Gamma_{\alpha\mu}N^{(\pm)}_{\alpha\mu}(-{\cal L}) a_{\mu} .
\eea
Elegantly, $N^{(\pm)}_{\alpha\mu}(-{\cal L})$
can be regarded as the counterpart of the Fermi function
$n^{(\pm)}_{\alpha}(-{\cal L})$ after accounting for
the level broadening.
Combining (\ref{A+-3}) with \Eqs{ME-3}-(\ref{rho-t}),
we complete the generalization of the formalism.

As an illustrative application of the generalized formalism,
we re-consider the transport through the (free) multi-level system.
Straightforwardly, the current expression of \Eq{IV-1} becomes
\bea\label{IV-2}
I &=& e \sum_{\mu}\frac{\Gamma_{L}(E_{\mu})\Gamma_{R}(E_{\mu})}
        {\Gamma_{L}(E_{\mu})+\Gamma_{R}(E_{\mu})}  \nl
  & & \times \int\frac{d\epsilon_k}{2\pi}
    \ti{a}_{\alpha\mu}(\epsilon_k-E_{\mu})[n_L(\epsilon_k)
    -n_R(\epsilon_k)]  .
\eea
This is the well-known formula for resonant tunneling
current, which is valid for arbitrary voltage
including the near-resonance situation.

\vspace{2ex} {\it Acknowledgments.} Support from the National
Natural Science Foundation of China, the Major State Basic
Research Project No.\ G001CB3095 of China, and the Research Grants
Council of the Hong Kong Government are gratefully acknowledged.


\begin{references}
\bibitem{note-1}
A more formal and rigorous treatment is possible
by improving the second-order Born approximation.
Namely, by self-consistently including higher order contribution
of the tunneling Hamiltonian $H'$, the operator
$\bar{a}_{\mu}(t)\equiv e^{\pm i{\cal L}t-\Gamma_{\mu}t}a_{\mu}$
would satisfy a master-type equation,
$\dot{\bar{a}}_{\mu}=\pm i{\cal L}\bar{a}_{\mu}-{\cal R}\bar{a}_{\mu}$,
and the second (damping) term coincides precisely with the one of
\Eq{rho-t}.
This treatment can be termed as a self-consistent Born approximation.
However, the simple implementation of a damping factor as done in present
Appendix will be more convenient in practice.
\end{references}
\end{document}